\documentclass[twocolumn,showpacs,preprintnumbers,amsmath,amssymb]{revtex4}

\usepackage{graphicx}
\usepackage{dcolumn}
\usepackage{bm}
\usepackage{psfig}

\begin{document}
\draft

\title{The evolutionary advantage of diploid sex}

\author{A.O. Sousa, S. Moss de Oliveira and J.S. S\'a Martins}
 
\address{\it Instituto de F\'{\i}sica, Universidade Federal 
Fluminense \\ Av. Litor\^anea s/n, Boa Viagem, 24210-340 Niter\'oi, RJ, 
Brazil}

\date{\today}

\begin{abstract}
We modify the Penna Model for biological aging,
which is based on the mutation-accumulation theory, in 
order to verify if there would be any evolutionary advantage of  
triploid over diploid organisms. We show that this is not the case, and 
that usual sex is always better than that involving three individuals.

\end{abstract}


\maketitle

\section{Introduction}

The reasons why sexual reproduction appeared, hundreds of million
years ago, are still under investigation today \cite{1}.
One of the reasons commonly used to justify sex is the higher genetic 
diversity it produces if compared to asexual reproduction. 
On the other hand, sex is closely related to aging: 
all sexual species present aging, which starts as soon as  
reproduction starts \cite{2}. 
Because of its ability to address both questions in a single context, 
the Penna model for biological aging \cite{3} has been extensively used 
in recent years to study why sex evolved (for a review see \cite{4}). 

S\'a Martins and Moss de Oliveira, for instance, have shown that the higher 
diversity of sexual populations can prevent extinction if a catastrophe 
damaging the genome of the individuals (such as exposure to radiation) 
occurs \cite{5}. 
Diversity was also the key issue discussed by S\'a Martins \cite{6} in his 
simulation of the competition of sexual and asexual varieties of a single 
species, coevolving under the assault of parasitic infestation. The density 
of parasites was shown to drive a transition between asexual and sexual 
orders, in agreement with the so-called Red Queen Hypothesis for the evolution 
of sex \cite{7}). 
The above mentioned results show that the model provides sexual varieties 
with survival advantages, due to their larger genetic diversity, over the 
asexual ones, whenever they compete in a changing environment. 
Nevertheless, in the absence of competition and a mutant environment, the 
asexual populations resulting from simulations of the standard Penna model 
are always much larger, since they generate twice as many offspring for the 
same population size. This effect is not overcome by the need, in diploid 
individuals, for homozygose if the harmful allele is to become active in 
loci for which it is not dominant. 
Without catastrophes or parasites, if a sexual and an asexual
population are allowed to evolve under the same immutable environment, 
competing for the same food resources, the asexual one will win and the 
sexual one disappears 
(due to the Verhulst logistic factor explained in section two).
Only very recently has a new ingredient been introduced \cite{8} into 
the Penna model allowing larger sexual populations to be obtained, with sizes 
comparable to the asexual ones. 

We now use this modified Penna model to address a tantalizing issue: since a 
diploid sexual population has the upper hand when competing against an 
asexual one due do the diversity generated by the use of genetic material 
coming from two different parents, why does not nature enhance this effect 
by allowing the genome of the offspring to benefit from three different 
templates? Is the need for homozygose mentioned above enough to overcome the 
burden of using three individuals to generate one offspring? To provide 
answers to these questions, we compare two different 
kinds of sexual populations: one involving the mating of two diploid 
individuals as parents (normal sexual reproduction) and the other involving 
three triploid individuals.
 
The paper has the following structure: In section 2 we describe the model, 
first for diploids and then for triploids, in section 3 we 
present our results and in section 4, our conclusions.

\section{The model} 
\subsection{Diploid population}
Each individual of the population is represented by a 
``chronological genome'', which consists of two bit-strings 
of 32 bits (32 loci or positions) each, that are read in parallel. 
One string contains the genetic information inherited from the mother 
and the other, from the father. Each position of the bit-strings is associated 
to a period of the individual's life, which means that each individual 
can live at most for 32 periods (``years''). Each step of the simulation 
corresponds to reading one new position of all individuals' genomes. 
Diseases are represented by bits 1. If an 
individual has two bits 1 in the $i$-th position of both bit-strings 
(homozygote), it will start to suffer the effects of a disease at 
his $i$-th year of life. If the individual is homozygous with 
two bits zero, no disease appears in that age. If the individual 
is heterozygous in that position, he will become sick  
only if that locus is one for which the harmful allele is dominant. The  
dominant loci are randomly chosen at the beginning of
the simulation and remain fixed.  
If the actual number of accumulated diseases reaches a 
threshold $T$, the individual dies.  

In order to avoid an exponential increase of the population and to introduce 
a dispute for food and space, a logistic Verhulst factor is used. 
Every time step and for each individual, a random number between zero and one 
is generated. 
This number is compared with $V=N(t)/N_{max}$, where $N(t)$ is 
the actual size of the population and $N_{max}$ is the carrying
capacity. If the random number is smaller than $V$, the individual 
dies, independent of its age or genome. For a discussion of other 
alternatives of implementation of this mean-field-like interaction 
between individuals of the population we direct the reader to Ref. 
\cite{cebrat}.

If a female succeeds in surviving until the minimum reproduction age
$R$, it generates $b$ offspring every period until death. The female
chooses randomly a male to mate, with age also greater than or equal to
$R$. The offspring genome is constructed from the parents' ones;
first, the strings of the mother are cut in two at a randomly chosen position 
(``crossing'') 
and two complementary pieces, one from each string, are recombined to 
generate the female gamete (one string of 32 bits). $M$ deleterious
mutations are then randomly
introduced. The same process occurs with the father's genome 
and the union of the two resulting gametes forms the
new genome. The sex of the baby is randomly chosen, with equal 
probability. Deleterious mutation means that if a bit 0 is 
randomly chosen in the parent's genome, it is set to 1 in the
offspring genome. However, if a bit already set to 1 is randomly chosen, 
it remains 1 in the offspring genome (no back mutations).

The description given above corresponds to the original sexual 
version of the Penna model \cite{4,5}. The new ingredient, mentioned in 
section 1 and introduced by S\'a Martins and Stauffer \cite{8}, consists 
in assuming that harmful mutation reduces the survival probability.   
Thus, at each iteration, or ``year,'' each individual survives with
probability $\exp (-m \epsilon)$ if it has a total of $m$ harmful 
mutations (taking into account dominant positions) in it's full genome 
(it is killed if a random number is tossed that is smaller than 
the survival probability). $\epsilon$ is a parameter of the simulation, 
fixed from the start. 
To summarize, an individual may now die for any one of three reasons: i)
randomly, due to the Verhulst logistic factor; ii) if its actual number 
of accumulated diseases reaches the limit $T$; iii) due to its survival 
probability being too small.

\subsection{Triploid Population}
In this case, we assume that mating involves three
triploid individuals (two males and one female or vice-versa). The 
chronological genomes consist of 
three bit-strings that are read in parallel. Homozygous positions 
are those with three equal bits at homologous loci. Harmful mutations are 
active only if there are three bits 1 at that same position, or at a 
heterozygous locus at which harmful mutations are dominant.
Only females generate offspring. There are random crossing and recombination 
to produce the offspring genome (see fig. 1), and deleterious mutations 
are randomly introduced in each of the three gametes. The baby is a
male or a female, with equal probability.

The first question we want to investigate relates to the competition between 
the benefit provided by a triploid genome and the effort involved in a 
mating that needs three individuals, instead of two, to generate the 
offspring. The benefit is the fact that, for triploids, mutations that happen  
in loci where the harmful allele is not dominant need to appear in all three 
bit-strings to become active.

\section{Results}

The curves presented below correspond to the average of the results
obtained for 20 different populations (20 different initial seeds for 
the random number generator), using the following parameters: 

\noindent Initial population = 10,000;

\noindent Maximum population size $N_{max} = 100,000$ individuals;

\noindent Maximum number of genetic diseases $T=3$;

\noindent Minimum reproduction age $R=8$; 

\noindent Birth rate $b=2$; 

\noindent Mutation rate $M=1$ per bit-string (or gamete);

\noindent Number of dominant positions $d=6$; 

\noindent Decrease in survival probability $\epsilon=0.015$;

\noindent Total number of Monte Carlo steps = 800,000; 

In figure 2 we present the time evolution of a diploid population 
(upper curve) and of two triploid populations, one for which reproduction 
involves the mating of one male and two females (central curve) and the other
corresponding to the mating of two males and one female (lower curve). 
In figure 3 we present the corresponding dimensionless survival rates $S(a)$ 
as a function of age $a$, $S(a)=N(a+1)/N(a)$, where $N(a+1)$ is the number of
individuals with age $a+1$ and $N(a)$ is the number of individuals with age 
$a$.

From these figures we can see that the diploid population is not only 
larger than the other two, but also presents a higher survival 
probability. However, it may be not enough to guarantee that the 
diploid population is better than the triploid ones. A second important 
measure, as mentioned in the introduction, is the genetic diversity of the 
populations. 
It has already been shown that the survival probability 
of a sexual population is the same as that of a diploid asexual population
that reproduces by meiotic parthenogenesis. However, the larger genetic
diversity of the sexual population may prevent it from
extinction if, for example, exposure to radiation occurs [5]. The
genetic diversity is calculated by measuring the Hamming distance, in this 
case defined by the number of different loci (bits) between the genomes, 
for all pairs of individuals. The probability distribution of these 
distances is obtained by making a histogram of the fraction of 
pairs, out of all possible pairs in the population, that present a given 
Hamming distance, normalized by its maximum possible value (64 for diploids
and 96 for triploids). Figure 4 shows the resulting distributions for 
the diploid and triploid populations. It is clear that the diploid population 
presents both a larger mean distance between pairs, indicated roughly by 
the position of the peak of the distribution, and a larger variance, 
measured by the width at half the maximum height of the curves. This assures 
a larger diversity within its genome space for the diploid populations, and 
will give them the upper hand when competing against the triploids 
under the pressure of a rapidly mutating environment.
It is worth while mentioning that the results are essentially the same if 
we allow a double crossing of the triploid genome during reproduction, and 
there is no benefit for the triploids to ensure that the offspring have 
their genetic material gathered from all three parents.

\section {Conclusions}

We generalized the sexual version of the Penna model for biological
aging to simulate also triploid populations. In these populations
individuals present three sets of homologous chromosomes. A harmful mutation
needs to appear in all of them to become active or to be in a locus where 
the harmful allele is dominant. We showed that normal diploid sexual 
populations have higher survival rates and are larger than the triploid 
ones, for the same carrying capacity of the environment. The genetic 
diversities of the populations, measured by the Hamming distance 
distribution, were also compared. The result shows that the diploid population
presents a higher genetic diversity than the triploid ones. We may thus 
conclude that, concerning survival probabilities, population
sizes and diversity, usual sex is better than that involving three
individuals. Sexual diploid populations would be favored in direct 
competition with triploids either in a stable or a mutating environment, 
and we claim that this is the reason for normal sex to have been chosen 
by evolution as the dominant reproduction strategy.  These results mean that 
the fact that triploids need mutations to appear in all three sets of 
homologous cromossomes to become effective is not enough to overcome the 
effort of a mating involving three individuals, and that more is not 
necessarily better insofar as genetic diversity is concerned.

\bigskip
\noindent Acknowledgments: We thank D. Stauffer and P.M.C. de Oliveira for 
a critical reading of the manuscript, and the Brazilian agencies CNPq,
FAPERJ and CAPES for partial financial support.

\newpage
\begin{figure*}
\centerline{\psfig{file=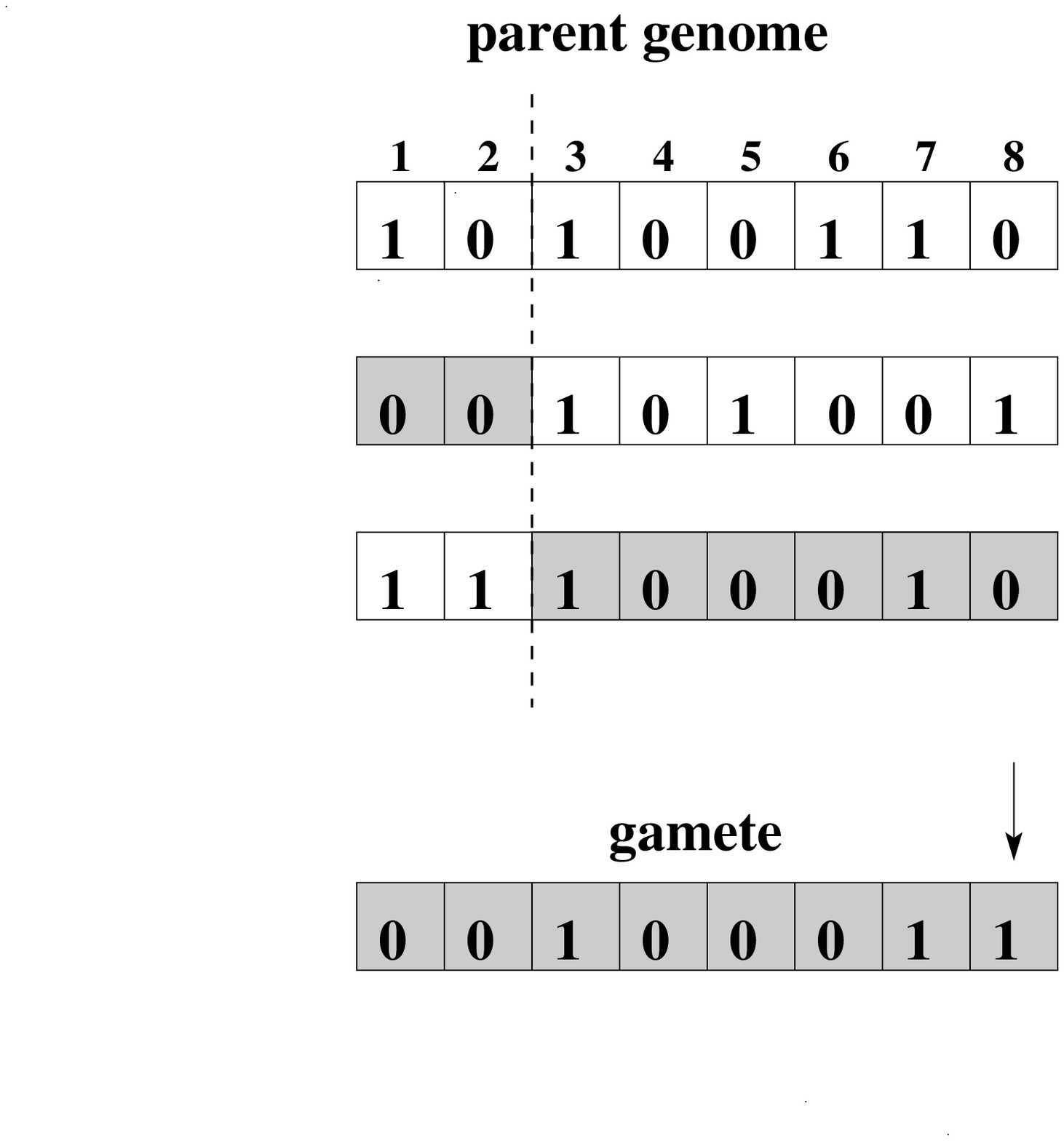,width=14cm,angle=270}}
\caption{\label{fig:crossing}Formation of one of the three gametes of a 
triploid individual. The dashed line indicates the position 
of the crossing, and the shadowed areas show the complementary pieces that 
will form the gamete. Both position and pieces are randomly chosen. The other 
two gametes are produced in the same way, from the genomes of the other two 
parents. The arrow indicates the position of a deleterious mutation added to 
that gamete. For diploid
populations, the scheme is the same, just omitting one of the bit-strings.}
\end{figure*}

\begin{figure*}
\centerline{\psfig{file=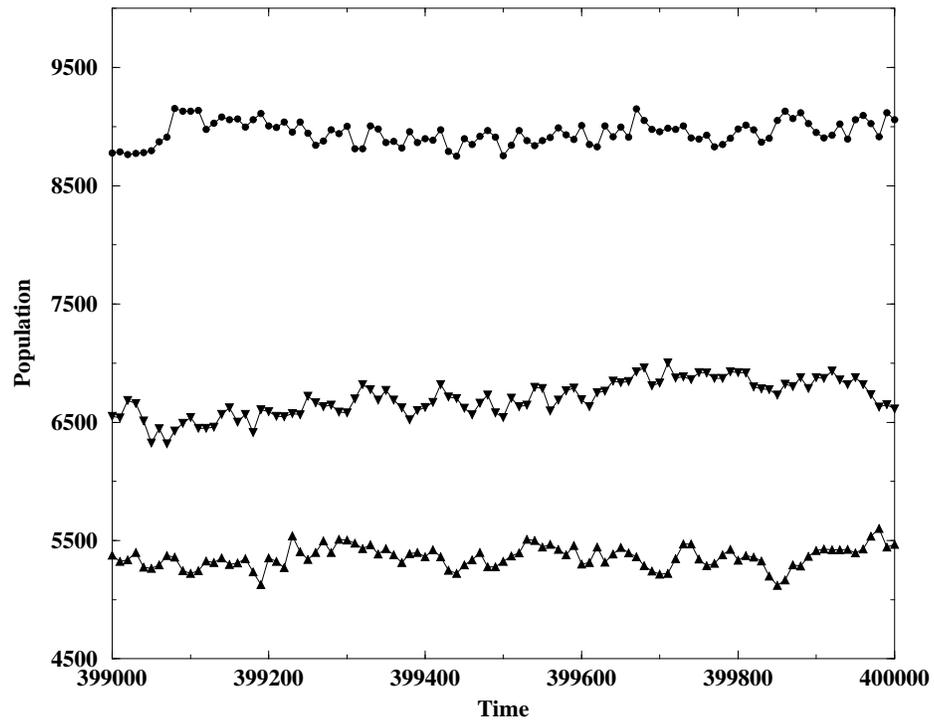,width=14cm,angle=-90}}
\caption{\label{fig:popul}Time evolution of a diploid population 
(upper curve) and two triploid populations: one for which reproduction 
involves one male and two females (central curve) and the other involving 
one female and two males (lower curve).}
\end{figure*}

\begin{figure*}
\centerline{\psfig{file=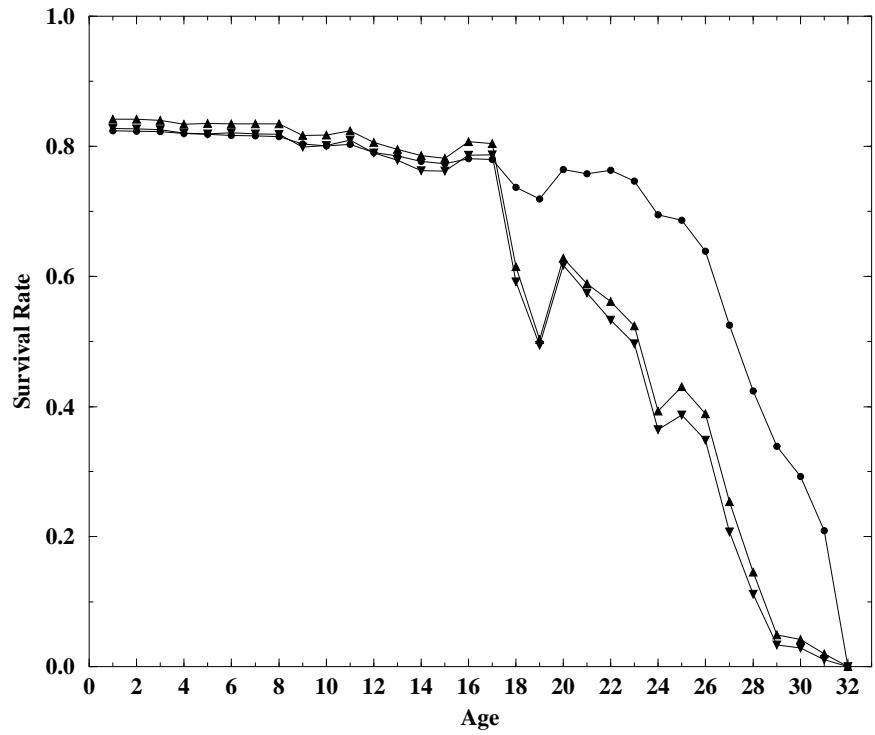,width=14cm,angle=-90}}
\caption{\label{fig:surv}Survival rates for a 
diploid population (full circles) and two triploid populations: one for which 
reproduction involves one male and two females (triangles down) and the other 
involving two males and one female (triangles up).} 
\end{figure*}

\begin{figure*}
\centerline{\psfig{file=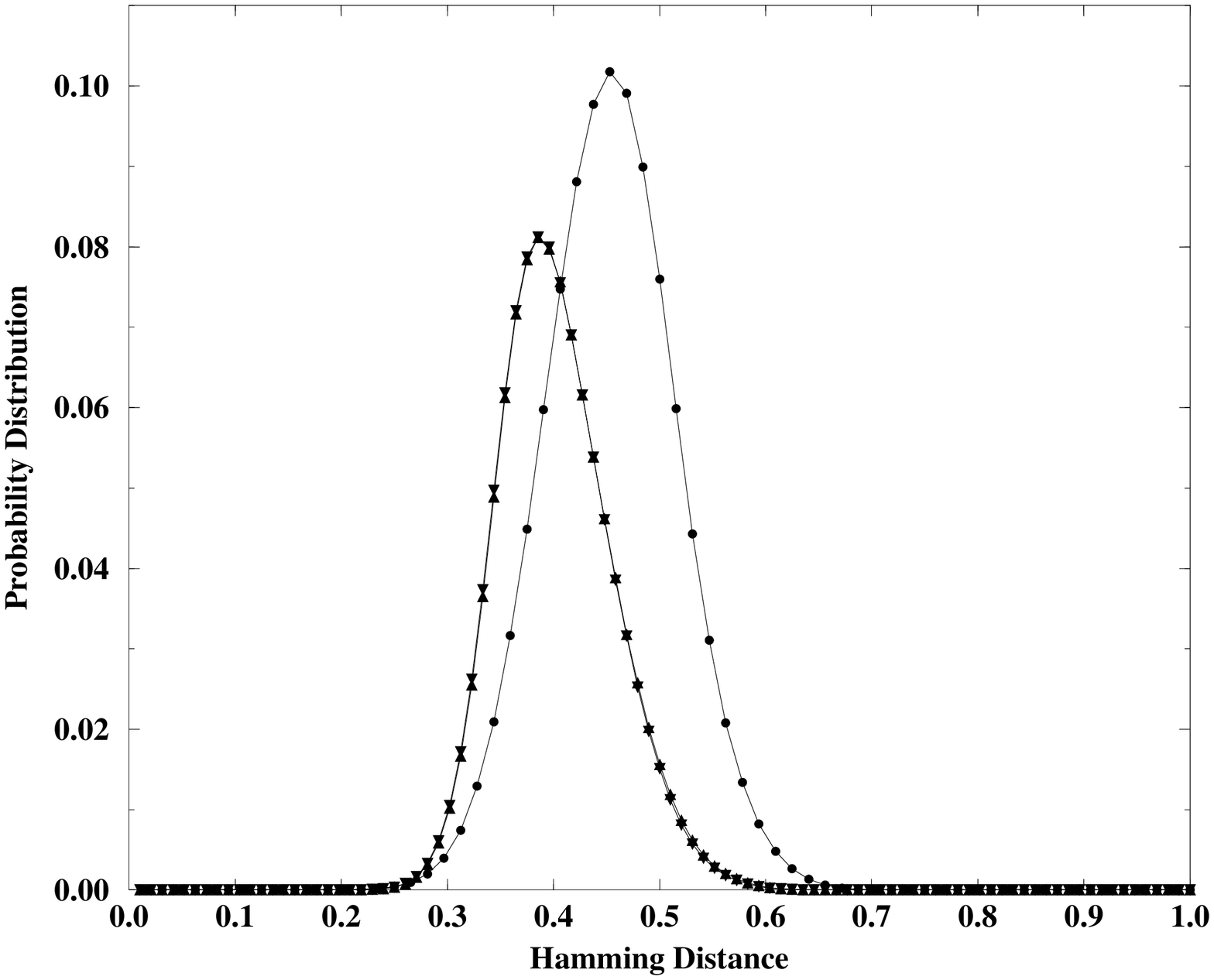,width=14cm,angle=-90}}
\caption{\label{fig:hami}Genetic diversity of a diplod population (full 
circles) and the two triploid populations mentioned in the captions of 
the previous figures, which are, for this particular measure, 
indistinguishable.
See text for details.} 
\end{figure*}


\bigskip
\begin{thebibliography}{99}
\bibitem{1} S. Dasgupta, Physica A 298 (2001) 465.
\bibitem{2} L. Partridge and N.H. Barton, Nature 362 (1993) 305.
\bibitem{3} T.J.P. Penna, J. Stat. Phys. 78 (1995) 1629.
\bibitem{4} S. Moss de Oliveira, P.M.C. de Oliveira and D. Stauffer, {\it
  Evolution, Money, War and Computers}, Teubner, Leipzig (1999).
\bibitem{5} J.S. S\'a Martins and S. Moss de Oliveira, Int. J. Mod. Phys. C 9
(1998) 421.
\bibitem{6} J.S. S\'a Martins, Phys. Rev. E 61 (2000) 2212.
\bibitem{7} B. Wuethrich, Science 281 (1998) 1982.
\bibitem{8} J.S. S\'a Martins and D. Stauffer, Physica A 294 (2001) 191.
\bibitem{cebrat} J.S. S\'a Martins and S. Cebrat, Theor. Biosc. 119 (2000) 156.
\end{thebibliography}
\end{document}